\documentclass[twocolumn]{svjour3}          % twocolumn
\smartqed  % flush right qed marks, e.g. at end of proof
\usepackage{graphicx}
\usepackage{cite}
\usepackage{amsmath,amssymb,amsfonts}
\usepackage{algorithmic}
\usepackage{graphicx}
\usepackage{graphics}
\usepackage{epsf,graphicx}
\usepackage{textcomp}
\usepackage{xcolor}
\usepackage{multirow, booktabs}
\usepackage{listings,multicol}
\usepackage{xfrac}
\usepackage[utf8]{inputenc}
\usepackage[T1]{fontenc}
\usepackage{fixltx2e}
\usepackage{csquotes}
\usepackage{amsmath,diagbox,tabu}
\usepackage{hyperref}
\usepackage{subcaption}
\usepackage{comment}

\title{An authentication protocol based on chaos and zero knowledge proof}

\author{Will Major \and William J Buchanan (Corresponding Author) \and Jawad Ahmad}

\institute{Blockpass ID Lab, Edinburgh Napier University, Edinburgh. UK.\\Corresponding author: William J Buchanan, w.buchanan@napier.ac.uk}

\date{29 Sept 2019}

\authorrunning{Will Major et al.}

\begin{document}

\maketitle

\begin{abstract}
Port Knocking is a method for authenticating clients through a closed stance firewall, and authorising their requested actions, enabling severs to offer services to authenticated clients, without opening ports on the firewall. Advances in port knocking have resulted in an increase of complexity in design, preventing port knocking solutions from realising their potential. This paper proposes a novel port knocking solution, named Crucible, which is a secure method of authentication, with high usability and features of stealth, allowing servers and services to remain hidden and protected. Crucible is a stateless solution, only requiring the client memorise a command, the server's IP and a chosen password. The solution is forwarded as a method for protecting servers against attacks ranging from port scans, to zero-day exploitation. To act as a random oracle for both client and server,  cryptographic hashes were generated through chaotic systems. 

%Experimental work focuses on prototyping port knocking implementations with the components introduced from the literature. A port knocking prototype is created, blending a zero knowledge proof of identity, and a chaos-based hash function, its successor (the following prototype) combines random beacons with a chaos-based hash function. Lastly, a prototype named Crucible is explored, combining BLAKE2 and Argon2 with the random beacon from previously, to form the proposed port knocking solution of this dissertation. Crucible is a stateless, secure port knocking prototype offering simplicity and usability over other port knocking implementations. Crucible is evaluated against a number of theoretical attacks, particularly against identification schemes, and found to have no large concerns. Suggestions for improvement on the previous designs are included, and these prototypes still offer advantages in different settings.\\

%Abstract: Problem; Aim; bit other work; implementation; significant result; a bit of a conclusion
\end{abstract}

\let\clearpage\relax
\section{Introduction}

Port knocking, if integrated into a security environment, can offer an additional layer of authentication for servers, furthering a defence-in-depth approach, and can conceal the presence of services. It is suited to defending against attacks directed at servers, ranging from automatic scanning, as part of attack-chain reconnaissance, to precisely targeted zero-day exploitation. Port knocking solutions have progressed and changed dramatically, since their conception as a simple tool for opening firewall ports. Many modern port knocking implementations have accumulated layers of complexity in the process of removing specific vulnerabilities from their predecessors. This complexity issue is further exacerbated by components in port knocking that are mutually incompatible: replay protection can result in desynchronisation problems, and interactive authentication requires the server to forgo its `silent' role. To combat this, many modern port knocking implementations in academic literature take an ad-hoc approach to fixing existing vulnerabilities without considering a holistic viewpoint, nor the minimalist lineage upon which port knocking was founded.

Port knocking can be considered a stealthy method of authentication and command execution, allowing a covert channel to exist between a client and server, across an untrusted network such as the Internet. When implemented properly, port knocking should be difficult to discover through passive surveillance of network traffic, or active reconnaissance of the server. Port knocking allows a server to conceal not only its individual services, but also its {\it role} as a server. Numbers Stations are a Cold War era covert channel using radio broadcasts of spoken number values (amongst other methods), suspected to communicate with intelligence assets in the field \cite{BBC}. In modern day computing, Meltdown \cite{melt} and Spectre \cite{spec22} are examples of serious vulnerabilities enabling covert channels for exfiltrating data from a victim's machine. 

Port knocking can simply be described as a means for communicating with a machine that is protected by a closed-stance firewall. This entails a client machine, operated by a user, sending a message -- or {\it knock} -- to a firewall. The firewall in this instance will be referred to as the {\it server}, for its role in receiving the knock, and the subsequently authorised actions it performs. This terminology further aims to clarify in situations involving multiple intermediaries (such as {\it additional} firewalls) between client and server. The networks across which this transaction can take place could be local, or, as is the common use case, remote. Port knocking is by design aimed at operating over untrusted networks, such as the Internet, wherein malicious actors of varying capabilities aim to subvert security measures. Port knocking methods generally use cryptographic primitives such as hash and encryption functions. In the proposed port knocking method, we have utilised the Chirkov standard map and absolute chaotic map for the hash that ensures the output knock is different for each session. Additionally, the chaos-based hash function provides a lightweight secure solution and as a result, the proposed scheme will not require third-party libraries.  Moreover, chaos-based port knocking scheme will have a number of properties such as sensitivity to initial conditions, non-periodicity, ergodicity, and attack complexity which strengthen the port knocking scheme.

This paper offers three novel port knocking prototypes: zero knowledge proofs and chaos-based cryptography; a combination of chaos-based cryptography and random beacons; and `Crucible' which is combines random beacons and password-based key derivation. Replay protection and NAT compatibility are considered significant issues in port knocking technology. This paper proposes novel approaches for dealing with these problems. 

The rest of paper is organised as follows. Related work is discussed in Section 2. Section 3 discusses design and the proposed methodology. Experimental results are presented in Section 4. The proposed method is evaluated in Section 5. Research findings are concluded in Section 6.

\section{Related Work}

Port knocking allows authentication to a host without requiring open ports, and without requiring modification of the underlying protocol \cite{1f}. Furthermore, the presence of port knocking can be difficult to detect by sniffing traffic, and almost impossible to detect by probing a server \cite{1d}. By making services `invisible', a machine's function as a {\it server} can be hidden from an attacker \cite{1t}, offering a level of anonymity. Port knocking disrupts attacker reconnaissance by denying fingerprinting efforts against open ports. This prevents automated and manual scanning efforts, and reduces ``malicious information gathering capabilities'' \cite{1h}. In the same vein, vulnerability scanning and discovery are impeded, and thus port knocking can be considered one of the few non-reactive defences against zero-day attacks \cite{1j}. In this manner, port knocking can provide protection for legacy and proprietary services with ``insufficient integrated security'', or for services with ``known unpatched vulnerabilities'' \cite{1v}. Port knocking also provides an additional layer of security and authentication, that any malicious actors must overcome before attacking the hidden service itself \cite{1v, 1h}.

Port knocking can incur load and performance loss on networks and systems \cite{1d}, the latter applying an overhead for each connection \cite{1f}. In addition, a number of ports may need to be ``allocated for exclusive use by port knocking'' \cite{1w}. Port knocking implementations may require user training \cite{1f} and client systems need to implement port knocking, which may require maintenance of dedicated remote client software \cite{1h}. Authentication in port knocking is largely facilitated through pre-shared keys (or other secrets), meaning secure key distribution and management become requirements. Port knocking also adds an additional layer of complexity into the protection of assets \cite{1h}, and another attack vector to defend against. The fail-closed stance of port knocking may result in inaccessibility of services, should the authentication mechanism fail on the server \cite{1w,1f}. Such a failure in this mechanism could render the server ``unreachable or more easily compromised'' \cite{1h}. 

\begin{comment}
If the traffic patterns of port knocking are readily identifiable then this can result in an attacker identifying client machines, and servers, alike \cite{1d}. Though not precisely a {\it drawback}, as mentioned earlier -- port knocking is not suitable for protecting open and publicly accessible services \cite{1f} (save perhaps a large overhaul in protocol frameworks). As a result, port knocking can't be implemented with ``with open access protocols such as HTTP/S, DNS and anonymous FTP which are among the top ten most popular services on the Internet'' \cite{1p}.
\end{comment}

% authentication association

\subsection{Mechanics and Architecture}\label{sec:mechanics}
One of the main differentials when considering port knocking solutions is the number of packets that are required to authenticate the client with the server. Traditional port knocking solutions sent a series of packets (each {\it knock}) across the Internet, where the {\it authenticating data}, would be communicated via this collection of knocks, known as a {\it knock sequence}. More recently, a single packet has been used to wholly and atomically send authenticating data to the server, in a single knock, giving rise to the term ``Single Packet Authorization'' \cite{1z}. \cite{1d} raises an immediate problem with using multiple packets to authenticate, in that network and routing issues between the server and the client, such as latency (from congestion \cite{1h}) or packet drops, will cause the authenticating packets to arrive out of order. If the knock sequence doesn't match the server's expectations, the client likely won't be authenticated. \cite{1y} offers a solution to this problem by including a sequence number within each knock's authenticating data, allowing the knocks to be reassembled after they are received. 
%\cite{1c} instead proposes introducing inter-packet delays (between each knock) to improve the likelihood of correct ordering, finding that this approach is effective at removing the problem, at the cost of tripling the knock sequence time. However, a denial of service vector is also heightened by this approach \cite{1d}: an attacker impersonating the client could disrupt the legitimate knock sequence with an invalid knock, denying the client authentication. \cite{1i} notes a further DoS scenario that could be created by an attacker impersonating a number of clients, with each spoofed IP initiating the knock sequence with the server, each requiring an allocated buffer, and therefore eventually exhausting server memory. The paper tests this theory, showing that a low end computer can be used to exhaust a much more powerful server.

\begin{comment}
In contrast, \cite{1z} highlights that using multiple packets may make the knocking appear similar to a port scan, which could go unnoticed, and offer an element of {\it stealth}. \cite{1z} adds that if port knocks are spaced out across traffic, this could make identification of port knocking difficult, from an attacker's perspective. In a similar vein, if a single packet is used (or a smaller sequence of knocks) an attacker will discover more about the protocol from each detection \cite{1d}, and an attacker may quickly correlate a single packet with the following authenticated action \cite{1z}. On the downside, sequences with a large amount of knocks can take up to a minute to execute \cite{1h}.
\end{comment}

\subsection{Non-Interactive or Interactive Server}\label{sec:interractive}
While traditionally communications in a port knocking implementation would be limited to unidirectional client to server messaging, some modern variations conversely include server to client communication. This could also be described as unilateral or bilateral communication. \cite{1a} for example, use client server conversations to negotiate a session key, which is then used to authenticate application data {\it after} port knocking has concluded. \cite{1c} includes a lengthy discussion on this topic. Incorporating challenge and response mechanisms into port knocking can be seen to enable {\it fresh} authentication, whereby random challenges are used to prove the identity of a peer. \cite{1y} also reiterates the advantages of interaction in providing freshness. This approach avoids the drawbacks of other solutions for replay protection, such as requiring time or state-based synchronisation, as is discussed further in Section \ref{sec:replay}.  \cite{1q} implement such a solution, where as proof of identity, the client is sent an encrypted random number, which it must decrypt and return, to prove it possesses the associated pre-shared key, thus authenticating the client. The authors describe this method as mutual authentication, though it is unclear exactly how the server is authenticated to the client. 
%\cite{1t} uses the TCP handshake as a challenge response mechanism, where the initial segment (i.e. knock) sent by the client contains authenticating data. This data is then validated by the server which in turn confirms success in its reply. \cite{1y} has the client provide a Message Authentication Code (MAC) of a server's provided random number, where the MAC includes further details of the knocking request, and is derived using the PSK. The authors note, if the server isn't authenticated to the {\it client} then an attacker can impersonate a port knocking server, and advance this into a full man-in-the-middle attack. To resolve this issue, the authors propose that the client could provide a random number, and have the server provide its MAC, thus providing mutual authentication, at the cost of an extra message. In contrast with these solutions, \cite{6i} succinctly states that only a single packet should move across the network.

\subsection{Multi-Party and Multi-Channel Involvement}
More recently, port knocking solutions have been forwarded that eschew the traditional client-server dynamic, and opt to include additional parties in the protocol. \cite{1d} sends the client cryptographic information, including a one time key, and a random number, via an out of band, dedicated SMS channel. The random number is used to generate a client-spoofed IP, for preventing client identification by a listening attacker. The one time key is used to encrypt the client's authenticating data, which is then sent as a port knock to the server. As each knock attempted in this way is unique, protection against replay attacks is provided, with the added bonus of making the port knocking traffic difficult for an attacker to identify. \cite{1n} uses a similar approach, where SMS is instead used to deliver information from which IPSec tunnel keys are derived, and used to setup a secure channel between the client and server. \cite{1f} enforce time synchronisation between client and server by having them issue NTP requests ahead of a knock, and on startup, respectively.
Maintaining accurate timing allows the time-based authentication to be more granular, reducing the window of opportunity for replay attacks. 
%In this scenario, the NTP server fulfilling the requests can be viewed as a trusted third party; if time is not synchronised in this setup, the authentication will fail. \cite{1p} offer a protocol for authenticating port knocking requests with keys established through DNS. As web services online are commonly accessed by name (e.g. URL) rather than by IP, the authors describe a system whereby a server's DNS TXT record is used to distribute a key, which is then included as part of the client knock. Their technique is aimed at defeating automated scans using IP-based targets.

% this section needs more criticism/security flaws
% SMS dangers - 
%   https://blog.vasco.com/authentication/sms-authentication/
%   https://pages.nist.gov/800-63-3/sp800-63b.html

%dependencies

\subsection{Replay Protection} \label{sec:replay}
A port knock authenticates the client to a server, typically by proving the client possesses a secret, such as a key or password. In proving this over the wire, encryption or hashing are employed to prevent eavesdroppers from learning the secret. An adversary listening to communications between the knocking client and listening server could record a knock and {\it replay} it back to the server in order to repeat its intended effect. Such attacks have been cited as the main vulnerability affecting basic port knocking implementations \cite{1h}. These attacks are prevented by ensuring that each time a port knock is executed, the authenticating data is unique, or {\it fresh}, and further ensuring that an attacker is unable to generate a fresh knock. Many of the techniques used to achieve freshness carry similarities with one time password (OTP) protocols \cite{1aa}.

\section{Design and Methodology}

\subsection{Zero Knowledge}\label{chap:ZKP}
A zero knowledge proof (ZKP) is a cryptographic protocol allowing one to prove they posses information to a verifying party, without revealing any of the underlying information itself \cite{2a}. Alternatively defined by \cite{2e}, a ZKP shows ``a statement to be true without revealing anything other than the veracity of the statement to be proven''.\\ 

As a real-world example (modified from \cite{2b}), to prove someone could distinguish between two different types of wine, a number of blind-tests could be performed until the claim was proven or refuted. If the claimant identified the wine correctly each time, the verifier can be reasonably sure of the claimant's distinguishing ability, without themselves learning the difference between the vintages. In a similar vein, if after being repeatedly being sent into the labyrinth as a sacrificial offering for the Minotaur, a single Athenian continued to escape, one could be reasonably certain the citizen knew how to escape, though one would not be able to discern the citizen's method. These are not perfect examples, but demonstrate the general idea.

\begin{comment}
Three properties are required of a zero knowledge proof, as described in \cite{2j}, these have far more complex mathematical ramifications, but are included here to give an idea of what constitutes a ZKP:
\begin{itemize}
    \item {\bf Completeness} A prover should be able to convince the verifier of a true statement with high probability.
    \item {\bf Soundness} The probability of convincing a verifier of a false statement should be low.
    \item {\bf Zero-Knowledge} Any information available after observing the interaction was already available beforehand.
\end{itemize}
\end{comment}

Zero knowledge proofs harness difficult mathematical problems to provide security against information leakage from the proof -- the `zero knowledge' property. These difficult mathematical problems can be seen in RSA, for large integer factorisation, and Diffie-Hellman, relating to the discrete logarithm problem. One-way (or {\it trapdoor}) functions are the embodiment of these mathematical problems, a function that is relatively easy to compute, but computationally infeasible to reverse \cite{2f}, also referred to as a quality of {\it intractability}.\\ %The Fiat-Shamir protocol uses modular squaring as its one-way function, in an environment where finding square root of a number is prohibitively difficult \cite{2j}.\\

Zero knowledge protocols are classically ``instances of {\it interactive proof systems}, wherein a prover and a verifier exchange multiple messages (challenges and responses)'' \cite{2l}, though there are a subset of ZKP where this interactivity is removed, reducing the number of rounds needed to establish and verify the proof. These `rounds' are akin to each wine tasting, or labyrinth escape, in the previous examples. A {\it non-interactive} zero knowledge proof (NIZKP) allows the prover to publish a proof, in a single communication, that can be openly verified by anyone \cite{2a}.\\

In the `proof' component of a ZKP, the prover asserts a verifiable claim. One such strategy is where a {\it proof of knowledge} is made by the prover, examples of this are seen in the previous review of port knocking mechanics: pre-image resistance of a cryptographic hash could imply proof of knowledge of the digested secret, proof of decryption could be considered proof of knowledge of the decryption key. Related is the concept of {\it proof of identity}, where a ``person's identity can be linked to his ability to do something and in particular to his ability to prove knowledge of some sort'' \cite{2b}. Identification is a natural application for proofs of knowledge \cite{2m} and as such, a proof of knowledge protocol can be used to authenticate \cite{2c}. For example, a user could prove they know a password, without revealing any information about the password itself, to any parties privy to the conversation. 

% string CRS vs random oracle

\subsubsection{Identification Protocols}
To harness the capabilities of zero knowledge proofs for the client (or {\it prover}) authenticating with the port knocking server, an {\it identification protocol} sets out the framework for what is required of each party, how values are calculated and which values are sent as messages. Identification protocols based around zero knowledge proofs are examined in depth in \cite{2a} and \cite{2l}, both works explore the protocols of Feige-Fiat-Shamir, Guillou-Quisquater and Schnorr:

% all 3: require randomness, support the inclusion of IDs?, setup of a secret

\begin{itemize}
    \item {\bf Feige-Fiat-Shamir} The protocol uses the difficulty of extracting square roots modulo a large composite integer \cite{2l}. FFS requires a variable number of rounds (depending on desired security level), each requiring 3 messages \cite{2a}. 
    \item {\bf Guillou-Quisquater} GQ uses the difficulty of extracting roots of a higher order \cite{2l}. GQ can be run in multiple rounds, or a single, and uses 3 messages per round \cite{2a}. GQ, at the cost of greater computation than FSS, minimises the number of interactions required, and is comprised of short, simple mathematical mechanics \cite{2a}.
    \item {\bf Schnorr} The Schnorr Identification Scheme harnesses the difficulty of computing discrete logs in a prime field \cite{2l}. Like GQ, Schnorr's solution uses simple mechanics (shown in Protocol 1, please see Appendix). The protocol uses 3 messages and one round \cite{2l}. Schnorr's has relatively low computational complexity and supports precomputation of some values \cite{2n}.
\end{itemize}

% cite these solutions
% give more history and comparison

The design goal of using only a single packet for authentication in the port knocking solution is at odds with the interactivity of the three identification protocols; specifically each requires (at least) three messages. \cite{2c} also notes ``a large number of interactions means a poor performance in both communication and in computation''. Thankfully, a solution to this problem is outlined in RFC 8235: ``Schnorr Non-interactive Zero-Knowledge Proof'' \cite{8235}. This RFC will form the reference basis for the remainder of this section.

\subsubsection{RFC 8235}\label{NIZKP}
The Fiat-Shamir Transformation is a method for transforming a three-message zero knowledge proof of identity into a single message equivalent protocol. This is achieved by introducing a cryptographic hash of particular variables, in lieu of the random challenge that is issued by the verifier (message 2). As outlined in the RFC, this can be used to convert Schnorr's scheme into a non-interactive zero knowledge proof, as seen in Protocol 2.

\subsubsection{Discussion}

Protocol 2 (please see Appendix) will be harnessed as a method for authentication, used during prototyping in the following subsection. Considering its actions as a black-box, where randomness, a private key and shared parameters are input, the result is a proof which will be sent over the wire to be received and validated by the port knocking server. The protocol requires a secure cryptographic hash and this will form the discussion in the following section.

% TBC
%   explain why this is difficult - DLP etc, look at serious crypto
%   sometimes known as signed schnorr? because the UserID and OtherInfo are signed - encyclopaedia pp 1123
%   recovery of A via public v? /  what does an attacker have access to?
%   why mod q?
%   what is DSA and how are parameters chosen
%   FSS transform/heuristic and the random oracle model
%   https://tex.stackexchange.com/questions/408900/how-can-make-a-protocol-in-latex
%   talk about modern use of NIZKP (examples from RFC) and applications such as blockchain
%   RE the public exponent checks - https://www.moderncrypto.org/mail-archive/curves/2018/000995.html

\subsection{Chaos}\label{chap:chaos}

\cite{3b} draws exact links between properties of chaotic and cryptographic systems. As seen in Table \ref{table:chaosvcrypto}, Shannon's descriptions of simple operations, of sensitivity to initial variable changes, and of a complex output, for strong cryptographic primitives, largely correspond with behaviours in chaotic systems. The process for mixing pastry dough used as a metaphor by Shannon and Hopf actually forms the basis of a well studied chaotic system, known as the Baker's map \cite{3h}.

\begin{table*}[htbp]
\tabulinesep=1.2mm
\small
    \centering
    \caption{Table comparing chaotic and cryptographic properties, modified from \cite{3b}.}
    \begin{tabu} to \textwidth {| X[2] | X[2] | X[2.1] |}
        \hline
\textbf{Chaotic Property} & \textbf{Cryptographic Property} & \textbf{Description} \\ \hline
Ergodicity & Confusion & The output has the same distribution for any input \\ \hline
Sensitivity to initial conditions/ control parameter & Diffusion from a small change in plaintext/key & A small input deviation can largely change the output\\ \hline
Mixing property & Diffusion from a single plaintext block & A small local deviation can largely change the whole space \\ \hline
Deterministic dynamics & Deterministic pseudo-randomness & A deterministic process can cause a random-like behaviour  \\ \hline
Structure complexity & Algorithmic complexity & A simple algorithm has a very high complexity  \\ \hline
\end{tabu}

% tbc check the table against [3c, pp.37] and the other source that included it, some of the wording is off

\label{table:chaosvcrypto}
\end{table*}

To help illustrate the key terms in Table \ref{table:chaosvcrypto}, the following definitions are included:
\begin{itemize}
    \item {\it diffusion}: the spreading out of the influence of the function input over many bits of the function output \cite{3d}. For example, a single plaintext bit being distributed over the ciphertext output.
    \item {\it confusion}: the use of transformations to obscure the the statistical dependencies between function input and output \cite{3d}
\end{itemize}

Lastly, it is important to clarify between some of the key terms introduced, particularly to delineate between the concepts of chaos and randomness. Truly random behaviour is non-deterministic, even if a random system is fully understood, predicting its outcome at a given future state is impossible \cite{li2018cryptanalyzing}. Chaotic behaviour, in comparison, is deterministic, and every future state in the system is determined by its prior initialisation \cite{li2018cryptanalyzing}. In applied cryptography, both {\it pseudo}-random and chaotic systems are deterministic, and used for their qualities of being {\it computationally unpredictable}, meaning that guessing the previous state of the system (e.g. the inputs to a cryptographic primitive) is computationally infeasible \cite{li2018cryptanalyzing,3d}. This is similar to how hard mathematical problems were used to hide secret information in zero knowledge proofs.

\subsubsection{Chaotic Systems}
Chaotic systems are often result from mathematical functions, or {\it maps}, such as the Ikeda map 
\begin{comment}
seen at the start of the section, in Figure \ref{ikeda}, is one such example. Another example is the 
\end{comment} 
and the Chirikov standard map. Mathematically, Chirikov standard map is written as:

%\begin{align}\label{eq1}
  %  p_{n+1} & = p_{n} + K \sin{\theta_{n}}  \mod{2\pi} \\
 %   \theta_{n+1} & = \theta_{n} + p_{n+1} \quad \mod{2\pi}\notag
%\end{align}

\begin{align}\label{eq1}
 \left\{
               \begin{array}{ll}
                p_{n+1} & = p_{n} + K \sin{\theta_{n}}  \mod{2\pi} \\
                 \theta_{n+1} & = \theta_{n} + p_{n+1} \quad \mod{2\pi}
               \end{array}
          \right.
\end{align}
where $K$ is control paramter, $p_n$ and $\theta_n$ are real values between $(0, 2\pi)$. Each increment of $n$ reflects an {\it iteration} of the map, much like how a hash function or ZKP uses a number of rounds \cite{3d}. The map represents a dynamical mechanical system, where the dimensions $\theta$ and $p$ are used practically to represent position and momentum, though the important note here is that variables take on new values, on each iteration of the map, resulting in a new $\theta p$-coordinate. The constant coefficient $K$ influences the degree of chaos exhibited by the map.
\subsubsection{Chaos-Based Cryptographic Hashes}\label{hashes}
Chaotic cryptography is an active research area, producing real-world applications including cryptographic primitives such as pseudorandom number generators (PRNG), encryption systems (both symmetric and asymmetric), and hash functions \cite{3f}. Chaotic maps and hash functions have similar characteristics \cite{3g}, as explored previously in Table \ref{table:chaosvcrypto}, and a number of cryptographic hash functions have been created harnessing chaotic systems.\\

For a formal definition, a hash function takes a {\it message} input, of arbitrary length, and calculates a fixed length output, known as the {hash value}, or a {\it digest}. This value is used to identify the input, acting as a fingerprint. There are a number of required properties for a hash function to be considered {\it cryptographically secure}, including:

\begin{itemize}
    \item {\it Preimage Resistance} - given a random hash value, an attacker should never be able to find the preimage associated with the value \cite{3c}. This is why hash functions are considered as {\it one-way} functions -- a message cannot be derived from a digest \cite{serious}.
    \item {\it Second-Preimage Resistance} - given a message $M_{1}$, with hash value $H(M_{1})$, an attacker should never be able to find $M_{2}$ such that $H(M_{1})=H(M_{2})$ \cite{3c}. 
    \item {\it Collision Resistance} - it should be computationally infeasible to find two (or more) different messages that hash to the same value \cite{3c}.
\end{itemize}

Returning to the components being researched for prototyping, a secure cryptographic hash function is required for use in Schnorr's NIZKP. From the design goals, this ideally will be lightweight, simple, and should not require the use of third-party libraries.

\subsubsection{Absolute-Value Chaotic Hash Function} \label{hash}
The chaotic hash function proposed by \cite{3k} forms the reference basis for the remainder of this section. This implementation of a chaos-based hash function was chosen for a number of reasons:
\begin{itemize}
    \item The chaotic map and design of the function is simple, using mostly basic mathematics, and the authors have provided pseudo-code of the algorithms involved.
    \item The authors have evaluated the hash using both chaotic metrics, and metrics required of secure cryptographic hashes. This includes:
    \begin{itemize}
        \item Assurance that sensitivity to initial conditions is reflected in the hashing process.
        \item Statistical analysis of confusion, diffusion.%, examining how many bits are changed in hash digests by flipping single bits in the message input.
        \item Comparative testing of the statistical analysis against other hash functions, including other proposed chaotic hash functions, and against popular hash functions MD5 and SHA-1.
    \end{itemize}
\end{itemize}

The hash function is {\it keyed}, meaning it falls into a category of cryptographic hashes that require a key {\it and} a message to produce the digest output. Like regular hashes, keyed variants can ensure the integrity of a message, though this is extended further to provide authentication (by use of a key) for the message. {In Ref \cite{luo2019ecm}, authors have utilised chaotic map i.e., Chebyshev for broadcast authentication and chaos-based hashing. All broadcast messages were authenticated via chaos maps.} A keyed hash has two main requirements to be cryptographically secure \cite{serious}: an attacker must not be able to {\it forge} a valid digest from a message without the key, and secondly an attacker must not be able to recover the key from message digests \cite{serious}. The hash function harnesses use of an absolute value chaotic map given by:

\begin{equation}\label{eq2}
    x_{n+1} = 1-ABS(\alpha x_{n})
\end{equation}

% explain abs
% explain range - monotonically decreasing for |X|=>2?
% values -1<xn<1?
% example values

\begin{figure*}[h]
\centering
\includegraphics[width=1\columnwidth]{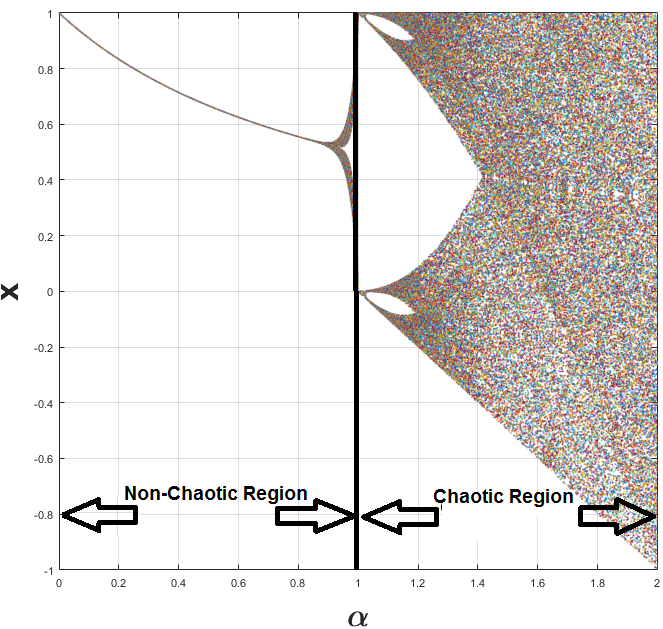}
\caption{{{Plot of Bifurcation diagram showing the chaotic region for $\alpha$ parameter.}}}
\label{alpha}
\end{figure*}

Equation \ref{eq2} produces a series of values that chaotically vary when $\alpha$ is chosen to reside in the interval $1< \alpha < 2$. {{The bifurcation diagram shown in Fig. \ref{alpha} highlights the range of $\alpha$ which can be used as a key parameter for chaotic region. From the Fig. \ref{alpha}, it is clear why one should use the interval $1< \alpha < 2$ when random data is required}}. The authors' provided pseudo-code outlines how this chaotic map is converted into a hash function, which is summarised in Protocol 3 (please see Appendix).

\subsubsection{Discussion}
To convert Schnorr's Identification Protocol (Protocol 1, please see Appendix) into a non-interactive zero knowledge proof, a hash function is required, to act as a random oracle for both client and server. Schnorr's NIZKP, in Protocol 2, now has its required hash function. Using only mathematical operations, this hash function should require few library imports, and therefore should be fairly platform agnostic. It's keyed mode of operation will be further used to authenticate messages from client to server.

\subsection{Random Beacons}\label{chap:beacon}
As discussed in Section \ref{sec:replay}, unique knock values are required to prevent an attacker re-sending a previous knock, to {\it replay} a previously authorised action. As port knocking solutions typically use cryptographic primitives such as hash and encryption functions, a random number is often added to the functions' inputs, ensuring that the output knock value is different for each session. This random value can be derived from methods such as iterative hashing, synchronised time clocks, or using counters. These methods require that both client {\it and} server have a way of reaching the same random value, i.e. it is {\it shared} between the two. Schnorr's NIZKP, as explored in Section \ref{NIZKP} does not suffer this problem: the random value is client generated, and included in each knock, and checked by the server on arrival. Setting Schnorr's NIZKP aside for now, the design goals explicitly prevent using client state, so an alternative replay prevention mechanism is desired.\\ 

\cite{4a} introduced the idea of a random beacon, an online security service emitting a random integer at regular intervals, for public consumption, with applications in cryptographic protocols requiring trusted, shared random number access. To introduce notation, the beacon {\it broadcasts} a {\it nonce} value (a number used once), to any satellite clients, who optionally may further process the broadcast through {\it extractor functions}. Formally, a list of requirements for a beacon service is provided by \cite{4f}:

\begin{itemize}
    \item {\it unpredictable}: an adversary should not be able to predict any information about the nonce prior to its broadcast
    \item {\it unbiased}: the nonce should be statistically close to uniform, random information
    \item {\it universally sampleable}: any satellite client should be able to harvest (or extract) the nonce
    \item {\it universally verifiable}: the beacon can be verified to be unknown to any party prior to its broadcast
\end{itemize}

% history
    % \cite{4e} has some
% applications of random beacons

% entropy definition

\subsubsection{Sources of Randomness}

A number of solutions are present in the literature for random beacons. \cite{4b} suggests that Network Time Protocol (NTP) can be used as a beacon. As an example, the hash of an NTP received time-stamp (within a safe margin of precision) could produce a pseudo-random nonce. This is similar to the port knocking implementation explored in Section \ref{sec:replay}, where clock synchronisation can be considered a shared state between client and server. Again, to avoid synchronisation issues between port knocking client and server, per the design goals, this is not an applicable solution for the prototype. Furthermore the time-measurements issued by an NTP service are likely unverifiable.\\

\cite{4c} propose using the price of publicly listed financial instruments, taken at closing time, as a source of random data. The authors note this approach has previously been used in cases including committee nomination, proof of work cryptographic puzzles, and in two public elections within North America. The proposed solution collects closing prices from a number of stocks and uses a extractor function (with elements of a PRNG) to harvest quantifiable entropy results from the financial beacons. \cite{4d} also provide an implementation using stock indices as random beacons. Their paper details considerations on particular stocks to harvest, which exchanges to use and associated timezone factors. Both papers note that there are historic examples of price manipulation in financial markets, a method by which the security of the random beacon could be adversely affected by a malicious actor. Such a scenario would constitute a breach of the unpredictability property of the random beacon. \cite{4d} suggest a greater problem may be posed by trusting the price reporting website (e.g. Bloomberg) to accurately reflect market values. \cite{4f} highlights that limited market opening hours will result in lack of beacon availability.\\

\cite{4e} discusses the newly introduced NIST random beacon, a public service providing a 512-bit nonce, every minute, of true-random data, generated using ``quantum mechanical phenomena''. This sounds ideally suited for the purpose, however, the authors note that while the service is extensively documented, there is no assurance to consumers that the numbers are generated as advertised. Simply put, again, the beacon is unverifiable. \cite{4f} treats NIST's reputation as a trusted party in this context with a heavy dose of scepticism, alluding to the discovery of a backdoor in the NIST-published Dual\_EC\_DRBG cryptographic standard.\label{sec:NSA}\\

Relating back to the beacon requirements, these solutions are all predominantly unverifiable, and as a result, can't be provably {\it unpredictable}, a desired quality. To address this problem, a number of different beacons can be used, where their results are combined or evaluated in a fashion that reduces the likelihood of tampering. Some of the methods used to tackle dishonest beacons include:

\begin{itemize}
    \item[] {\bf XOR nonces }\cite{4g}: in this solution each beacon service is polled, and the resulting nonces are XOR'ed together. If the beacon sources are incapable of influencing each other (perhaps they can't witness each other's broadcasts), then this method does reduce the effectiveness of tampering, by making the outcome more unpredictable. However, if the beacons {\it are} capable of influencing each other, the authors note that the last beacon to be polled could calculate the penultimate XOR result and alter its broadcast accordingly to influence the final outcome. In such a scenario, the solution is ineffective and redundant. 
    
    \item[] {\bf Round-robin nonces }\cite{4g}: this solution (also discussed in \cite{1ah}) simply rotates which sources are used by using each beacon for a fixed time-period, before moving onto the next. Unpredicability of the outcome is proportional to the ratio of dishonest beacons in the pool, meaning if at least one beacon is honest, then the round-robin approach does indeed makes the overall outcome more unpredictable. Unfortunately, this would require a shared state between the client and server, namely the position of the round, disagreeing with the design goals.
    
    \item[] {\bf Hashing nonces }\cite{4g}\label{sec:hashnonces}: the collection of nonces from different beacons could be hashed in some manner, the authors suggest that a hash of the concatenation of each nonce could be taken. The authors then argue, an adversary with a large amount of computing power at their disposal could use brute-force techniques to find preimages of the hash function with certain qualities (e.g. $H(m)$ starts with '00') resulting in the compromise of the other nonces in the pool. For these reasons, hashing nonces is considered by the authors as equal or less secure than the XOR strategy.
    
    \item[] {\bf Delayed evaluation }\cite{4h}: as discussed previously, a dishonest beacon can alter its nonce in an attempt to influence the other nonces it is combined with, to try and influence the final outcome. Variable delay functions (VDF) are processes designed to strategically {\it slow down} calculation of some task, and hence can be used to slow down the process of evaluating the pool of nonces. If for example, two independent beacons publish nonces in an hourly window, the delay function could be set to prolong the time it takes to combine these nonces. If the delay was set to a long enough period, neither beacon would have the time to derive a malicious nonce, before the hourly window had transpired and new nonce values were expected to be broadcast. \cite{4e} contains further discussion on VDFs and their applications.
    
\end{itemize}

% change 'evaluating' to extractor function /extraction

% TBC can use XOR? rand.org XOR NIST XOR BTC
% https://crypto.stackexchange.com/questions/48145/xor-a-set-of-random-numbers
% % TBC https://crypto.stackexchange.com/questions/17658/mixing-entropy-sources-by-xor
%       are the sources independent if 

These anti-tampering methods provide options for combining multiple random beacons, where the issue with trusting an single-source provider of random data appears to largely concern how predictability can be avoided, given that solutions are largely unverifiable, and could be tampered with. The proceeding section covers the random beacon chosen for the prototype, that is transparent, distributed, an accordingly conveys a higher degree of trust. It is worth noting however, that preferably if the literature had provided a suitable alternative, applicable to the design goals for the prototype, the combination of {\it multiple} random beacons would be the preferred option, as discussed in the evaluation. 

% problems of manipulation/ malleability and trust/lack of assurance

\subsubsection{Bitcoin Random Beacon} \label{BTC}
Blockchains are distributed, immutable ledgers of transactions used to record and authenticate activities of involved parties. Bitcoin is one such instance of blockchain technology known as a {\it cryptocurrency}, offering openly decentralised, self-governed currency transactions, backed by secure cryptographic protocols. Bitcoin transactions are collected into blocks, where within each block a Merkle Root is calculated and recorded in the block header. The Merkle Root involves hashing each transaction in the block in a manner that ensures that the entire contents of the block, including its ordering, is recorded to protect the integrity of the block's contents. Once a block is filled with transactions, the hash of the {\it previous block} on the chain is added into its header, forming the `chain' of collections of transactions, where each link ensures the integrity of its predecessor. \cite{4i} \\

The job of calculating the latest block on the chain is tasked to miners, who are rewarded with Bitcoin currency for their efforts. Alongside the aforementioned hashing calculations, miners must also solve a proof of work puzzle for each new block -- the block's hash digest must begin with a number of zeroes -- a task requiring millions of hash calculations. If multiple chains are broadcast to the Bitcoin network the longest is chosen, meaning it has been created with the most work, and therefore is the {\it consensus} reached by the mining community. This ensures that in order to tamper with the blockchain, an attacker would require computational power surpassing the rest of the mining community. \cite{4i}\\

The chosen random beacon for the prototype is taken from \cite{4f}, which forms the reference source for the remainder of this section. The paper proposes that the random values resulting from proof of work computations in the Bitcoin network make each new block mined on the network a suitable random beacon (please see Protocol 4, Appendix).

\begin{comment}
\subsubsection{Discussion}
Protocol 4's implementation of a random beacon should be secure, provided the HTTPS protocol is, and provided the data from the API is accurate. Using a keyed hash to digest the received block data ensures that the resulting random beacon value is {\it only} known to those in possession of the key, this is a welcome corollary for the beacon's use in general, as some protocols necessitate the use of a {\it private random} value. 
\end{comment}

% HMAC vs MAC vs keyed hash - is my chaotic hash a PRF?
%   see [serious]
%   https://crypto.stackexchange.com/questions/6523/what-is-the-difference-between-mac-and-hmac

% suitable as a public random beacon, first in literature requiring no trusted parties
% new blocks approx every 10 mins - poisson
% similar to the finance one
\begin{comment}
\subsection{Conclusions}
This chapter has introduced three key topics that will inform decisions in the following chapter, when prototyping to produce a secure port knocking solution. Schnorr's NIZKP provides an authentication protocol. Chaos-based cryptography has provided a keyed-hash function cryptographic primitive - a tool with ubiquitous use potential, like a Swiss Army Knife. Random beacons will ensure each knock is unique without using the replay-prevention methods seen in Section \ref{sec:replay} to be fraught with problems. These topics will form building blocks upon which new port-knocking designs are built.
\end{comment}

\section{Experimentation}
\begin{comment}
After direction from an extensive literary review, and an additional examination of modern and theoretical topics in cryptography, the ideas of the dissertation up to this point are synthesised in this chapter, through a process of prototyping. Three prototypes are exhibited in succession, each of which is designed to remedy the weaknesses of the previous iteration. The final prototype surviving the process will be selected as the proposed solution for this dissertation.

% developed in Python, platform agnostic, not finished products...
\end{comment}

\subsection{Prototype I: ZKP and Chaos}

The first prototype introduced here combines the work of Schnorr's Non-Interactive Zero Knowledge Proof (NIZKP) and the Chaos-based Keyed Hash Function, from \ref{chap:ZKP} and \ref{chap:chaos} in the preceding subsection. Prototype I will display a large amount of the general structure ahead of its successors.

\subsubsection{Setup}
The setup of a client and server is the only out-of-band process, used to derive the secrets and configurations required for port knocking operations. The subroutine achieving this, produces a {\it profile} JSON file each for client and server, housing cryptograhic parameters and configuration settings. The profile contents include:
\begin{itemize}
    \item Schnorr NIZKP parameters (refer to Section \ref{NIZKP})
    \begin{itemize}
        \item Group parameters $p,q,g$ are generated using a call to the \texttt{openssl} library. Specifically this uses the \texttt{dsaparam} module, to generate DSA parameters, which are applicable for the NIZKP as outlined in \cite{8235}. DSA key-length is set to 2048-bit.
        \item Private key $a$ is randomly selected from the range $[0,q-1]$ and is used to derive public key $A \equiv g^{a}$. Note: the only difference between the client and server profiles generated is the absence of $a$ in the server profile, as the ZKP proves knowledge of $a$.
        \item 8-bit command ID - for future development to support multiple commands under a single profile. In the original protocol, these were refered to as {\it UserID}, and {\it OtherInfo}.
    \end{itemize}
    \item Private hash key: a float number in the range $(-1,1)$, with 256-bit precision, for an associated 256-bit keyspace. Implemented using the \texttt{mpmath} 3rd party Python library.
    \item Server port number: randomly generated during setup, and fixed throughout all exchanges.
    \item Command: user specified shell command to run upon successful client authentication.
\end{itemize}

On a modern laptop, generation of the profiles was near instantaneous, and each was sized at 3KB. The profiles should be securely transferred and housed on the client and server machines.

\subsubsection{Chaos-based Hash Function}
The hash digest length was increased to 256-bit, as was required by RFC 8235 \cite{8235} for compatability with the NIZKP: ``The bit length of the hash output should be at least equal to that of the order q of the considered subgroup.'' This level of precision necessitated that the \texttt{mpmath} library be used, above the standard float data types that ship with Python. The horsepower required to manage calculations with these floats, with up to {{$\log(2^{256})/ log(10) \approx 78$}} decimal places, had a dramatic affect on the hashing speeds, with the hash taking over 20 seconds to calculate the required digest for the NIZKP protocol. Issues with chaotic map computations of floating point numbers were noted by \cite{3g}, though were not expected to be this severe. A large factor in the sluggishness of the hash function was the number of iterations to perform, or in terms of the chaotic map, the time parameter $t$. From the authors' paper, little guidance was set for how to select this parameter, $t=10000$ was one such baseline used for testing, however this proved unachievable in practical time. A further factor in this speed is undoubtedly increasing the hash's bit length. Lastly, as the hash was implemented in Python, a high-level language, time savings could further be gained by implementing the final version of the port knocking application in a language closer to the client hardware.

\subsubsection{Client Actions}
In Schnorr's NIZKP, a hash is taken of the public key $A$, the random walk $V \equiv g^{v}$, where $v$ is randomly generated by the client, and other, non-essential parameters. The resulting digest $c$, accompanied by $r \equiv v-ac$, form the proof of knowledge, or knock in this case, as the pair of concatenated values $c||r$, that are then transmitted to the port knocking server in the payload of a UDP datagram.\\

UDP was chosen as the transport protocol for sending the values, as TCP connection-based overhead and interactivity were not deemed necessary nor applicable, respectively, to minimalist design goals. ICMP traffic could be an alternative, though nothing was found in the literature to support this approach, whereas \cite{1y}, \cite{1g}, \cite{1r} provide justifications for choosing UDP. Python's native \texttt{socket} library was used to send the knock packet, where the server IP and destination port values are both taken from the client profile, the former being user input from \texttt{stdin}, and the latter being randomly generated.

\subsubsection{Server Actions}
The server parses traffic using \texttt{scapy}, a 3rd party Python library that provides an interface to lower-level \texttt{libpcap} functionality. The open-source \texttt{libpcap} library for network traffic capture operates on Linux systems, allowing packets to be inspected before firewall rules \cite{1h}, and is ``among the most widely used [APIs] for network packet capture''. Scapy is used to continually sniff the server's network interface. Collected traffic is then subjected to conditional rulesets designed to filter out traffic not meeting ZKP criteria: 

\begin{enumerate}
    \item Traffic must be destined for the server port, as setup in the client profile.
    \item Traffic must be UDP, with an integer payload.
    \item The payload is of length equal to the hash digest length, plus an integer $r \mod{q}$, padded with zeroes to fix this length.
\end{enumerate}

If a packet is sniffed meeting these criteria, then the Schnorr NIZKP Protocol (see \ref{NIZKP}) outlines the mathematical checks performed to validate the proof, computationally, this involves:

\begin{enumerate}
    \item Calculate $c$, $v$, by splitting the packet payload into their appropriate lengths.
    \item Calculate $V \equiv g^r  A^c$ using values from Step 1, and the client profile.
    \item Check whether the received $c$ is equal to:\\
    $\text{H} \left(V \| A \| \text{CommandID} \right)$
\end{enumerate}

If the proof is successful, the client is authenticated, and their requested shell command is executed. Though not implemented, the protocol allows for accompanying the proof with further data, this could be used to facilitate multiple command options for a single client, and multiple users.

\subsubsection{Discussion}
Prototype I, uses third party libraries only for floating-point computations, and packet sniffing from the wire. Theoretically, a powerful pocket calculator could generate the proof required to authenticate, and it could be sent via any number of on-line services for testing network connectivity. It could be argued that reducing dependencies such as 3rd party libraries use will reduce threat vectors from exploitation of those libraries. The prototype is further well suited for application on devices with limited ability to maintain updated libraries. Replay protection is a corollary of mixing randomly chosen $v$, and resulting $V$, in some form, into $c$ and $r$, ensuring each time a knock is sent, each element in the proof is derived from a nonce.\\

% The lengthy runtime of the chaos-based hash is a detractor, however the parallel design of operations ensures server availability is unaffected.

Returning to the design goals, the aim of having the server precompute acceptable knock values is entirely missed. Instead, this prototype requires the server to check a potential client proof, requiring two exponentiation operations in the group setting \cite{8235}, and a hash function execution. This opens a serious attack vector: by sending data to a suspected port knocking service port, an attacker can force the server to perform computations, potentially blocking out valid client knocks. As per \cite{1p}, ``processing required to verify the [knock] should be minimal and not introduce a [DoS] vector.''. Filtering options are limited for preventing such attacks, as knock values are pseudo-random, possessing no identifiable characteristics other than length, contradicting goals from \cite{1p}: ``unauthorized packets should be rejected as early as possible to reduce attack surface and decrease server side processing.''\\

To combat this, a traffic rate limiter could be enforced (the \texttt{scapy} sniffer offers such a feature) though this could in turn enable an attacker to purposely trigger the rate limiter to block out legitimate clients. Alternatively, the server's sniffer could establish a sort of reputational filter, ignoring the traffic from IPs that have sent invalid knocks, though this could be circumvented if the attacker spoofed their IP. Hopefully these scenarios illustrate why both the {\it only precomputation} and {\it only key-based authentication} were included as design goals. Suffice to say, this prototype does not fare well against denial of service attacks. DoS mitigation could be supplied by network and host monitoring solutions, such as IDPS, though this should not be required of a port knocking implementation.

% TBC do further comparison with the design goals
% c_hash is too processing expensive for X devices?
% difference between small/embedded/iot devices for client roles vs server roles
% better suited for enterprise roles?

\subsection{Prototype II: Chaos and Random Beacons}
The second prototype explored in this subsection forgoes zero knowledge proofs, and instead adopts a random beacon service, as described in \ref{chap:beacon}. The chaos-based hash functionality is further retained, though this time it is used to parse the random beacon. The client and server profiles are setup as previously, though without the parameters for Schnorr's NIZKP. Largely, the code is replicated from the previous prototype, with modifications outlined in the following.

\subsubsection{Blockchain-based Random Beacon}
The Bitcoin random beacon protocol, as shown in Section \ref{BTC}, pulls information from a website providing updates on Bitcoin's blockchain. The Python code for implementing this feature is taken from \cite{4j}, with small changes, including a change of website to the blockchain API provided by \cite{api}, which for this purpose requires only a single HTTPS GET request using the Python native \texttt{request} library, to a URL of the website's blockchain API. This API replies, offering information on the latest published block in JSON format. From there, the block header, and block header hash, are extracted and combined through a pairwise \texttt{OR}. The resulting string is then passed through the chaos-based keyed hash function, resulting in a 256-bit beacon value. Of note, if the hash function used were {\it not} keyed, then the attacker would be able to recreate the beacon output. Given that the key used in the hash function is only available to client and server, as per Protocol I's setup, this means the beacon values are shared, random and private. The {\it keyed} hash function could be used to combine multiple random beacons, for greater security (see Section \ref{sec:hashnonces}, though this is beyond the scope of the demonstration here.\\

After a new beacon has just been received, the beacon service sleeps for a fixed number of minutes, which can be changed to preference in accordance with block production speeds varying roughly around the 1 per 10 minutes mark \cite{4f}. Following this wait, the beacon will check more frequently, until the data pulled from the API differs, signalling a recalculation of the knock to expect from a client.

\subsubsection{Client Operations}
For this prototype, the UDP knock payload is defined as $H_{k}(\text{beacon}||\text{command})$, where $k$ and $H$ are the key and chaos-based hash implemented in the previous prototype. As discussed in the port knocking mechanisms literature review, hashes are one way functions that can provide a proof of identity: the attacker can only generate this payload if they are privy to the secret key, which is securely shared in out of band setup. Replay protection in this instance is provided by the freshness derived from the beacon. After processing, the beacon's values are pseudo-random, so the only instance in which the hash function receives two identical messages would be where the blockchain API reports an identical block and block hash. Therefore, the only instance in which a knock-collision occurs would result from either this scenario, or a weakness in the hash function. The security level resulting from a 256-bit keyspace for the chaos-based hash function ensures this is unlikely. Once constructed, the UDP payload is sent out as previously.

\subsubsection{Server Operations}
The port knocking server's operations include periodically harvesting random beacon data, converting this data into the knock it expects to see from the client, and monitoring the network to detect whether this knock has been received. These responsibilities require a degree of concurrency, for example the server must retain its listening abilities whilst at the same time calculating what the proceeding knock should look like, which uses the overwhelmingly slow chaos-based hash function. To handle these tasks in parallel, the native Python \texttt{multiprocessing} library is used to setup individual processes for traffic monitoring, beacon harvesting, and authorised command handling. Special variables handled by a \texttt{multiprocessing Manager} are shared between the processes, for instance to signal that the client has successfully authenticated. 

\subsubsection{Discussion}
Once a beacon has been harvested and processed, the server is left with a string value to listen for on the network, which if found, signals authentication, and subsequently the user's command is authorised. To enable multiple commands, a new $H_{k}(\text{beacon}||\text{command})$ is calculated for each, per beacon, and accordingly listened for. In comparison with the first protocol, and in the context of the design goals, this variant is vastly more simple, and has managed to eschew the denial of service attack vectors the latter was vulnerable to. Where previously the server performed exponential calculations and hashing operations to verify the authenticating data, this prototype need only check whether strings are equal. By removing the Schnorr framework, only a single private key is required (for the keyed hash), and the \texttt{openssl} library can be forgone. The introduction of a trusted third party (the blockchain API), is an obvious security detractor. Further examination of this aspect will follow later, in the evaluation section.

% applications
% length of data transmitted is less (c vs. c+r)
%   could mean better options for embedding
% on the other hand, if the server is compromised the client key is too, NIZKP didn't suffer that fault
% embedded devices, hard to update...if the server was compromised a bunch of retail products need recalled

\subsection{Prototype III: Crucible}

The final prototype introduced by this paper differs largely from the previous iterations. As a consequence of research leading up to this section, Prototype III, henceforth named Crucible (a namesake derived from its blending of ideas, much like elements in a furnace) draws its motivations from zero knowledge proofs, and chaos-based cryptography, but instead replaces both previously implemented components with dedicated off-the-shelf, cryptographic alternatives, that are already established in practice. \\

For authentication purposes, the previous review of zero knowledge proofs aimed to lead development towards a practical proof of identity scheme which could be used to prove knowledge of a secret key. Crucible pursues this line of thinking, but instead uses a password based key derivation function (PBKDF) to prove knowledge of a {\it password}. In doing so, the server does not possess the password itself, only its hash. In both previous prototypes, the run-time of the chaos-based keyed hash severely dampened results, instead, in Crucible, it will be replaced with a modern keyed hash alternative.

% aims
%   stateless
% chaotic hash lengthy runtime + discussion on variable delay functions -> tie in with PBKDF
% ZKP -> tie in with proof of identity, from proof of knowledge of the password, proving knowledge of the key
% 

\subsubsection{Password Based Key Derivation}\label{sec:Argon}
A key derivation function converts a secret value (such as a master key, passphrase, or password) into a secure cryptographic key \cite{serious}. Password based key derivation functions (PBKDF) are the family of such solutions aimed at converting potentially weak user supplied passwords into a keys that are specifically designed to be resistant to cracking attempts, making them well suited for storing user credentials on a server, whilst limiting the exposure to users if that server were to be compromised \cite{5b}.\\

As from the discussion on what comprises a secure cryptographic hash in Section \ref{hashes}, a PBKDF requires all the qualities of preimage, second-preimage, and collision resistance. In addition to this, a PBKDF needs resistance to lookup table attacks (e.g. rainbow tables), CPU-optimised cracking, hardware-optimised cracking (e.g. GPUs, FPGAs and ASICs), amongst other criteria, as established by the 2015 Password Hashing Competition \cite{5b}. Argon2 was the winner of this competition, and is selected for purpose in Crucible. Further details on Argon2 can be found in the IETF draft \cite{5c} and design paper \cite{5d}.\\

The prototype uses Python's native \texttt{passlib} library, updated with installation of the \texttt{Argon2} package. This specifically uses the Argon2i variation, recommended for password hashing purposes, as opposed to other KDF operations \cite{5c}. The hash parameters chosen for this prototype set Argon2 to use 20 rounds, 256MB ram, and 2 threads of parallelism, producing a hash in under 2 seconds on a modern laptop. Salt is set to a fixed value (explained later in Section \ref{sec:passwords}), and with the chosen parameters is stored alongside the digest, in a single hash string. These settings should be chosen in implementation to maximise computation efforts under the used hardware environment.\\

The setup process is similar to the preceding prototypes, where instead Argon2 replaces key generation procedures: the user is asked to input their chosen password (which they are to memorise), the hash of which is then stored only on the port knocking server. In addition to memorising the password, with Crucible the user needs to know only the IP address on which to knock, and the command name to execute. In this manner, Crucible is {\it stateless}, whereby a user can download the client application, and perform port knocking, without needing access to secret keys or other parameters, i.e. following installation of a profile on the server, a client profile (as per previous prototypes) is not required.

\subsubsection{Keyed Hash}
With the client able to derive a key from their chosen password, Crucible follows Prototype II in using a keyed hash to digest a random beacon. In preference of the chaos-based keyed hash already explored previously, an established hash function is taken from the literature. 

BLAKE2 was chosen to perform the task of hashing operations in Crucible, being among the fastest secure hash functions available, and the most-popular non-NIST standard hash \cite{serious}. Unlike alternatives Siphash, and SHA3, BLAKE2 is resistant to side-channel attacks, where an attacker has access to RAM and registers \cite{serious}. Further, unlike SHA3, BLAKE2 has native support for keyed mode hashing, without additional construction. More information on BLAKE2 can be found in \cite{5f}.

The particular variant BLAKE2b, was chosen for 64-bit optimisation in the test environment. The \texttt{pyblake2} 3rd party Python library \cite{5g} was used for the Python 2.7 implementation (referenced on the authors' website \cite{5h}), however the native \texttt{hashlib} supports BLAKE2 for later versions.

\subsubsection{Setup}
Ahead of port knocking operations, the client must generate a profile to leave with the server, used by the server to know which knocks to listen for. The generation process firstly prompts the user for a password, and the user is then directed to submit a number of commands for the server to execute on successful authentication. Each command must be named, and along with the password and IP of the server, each command name must be memorised. Following this, per the previous prototypes, the client profile is serialised in JSON and saved in a text file, to remain with the server.

\subsubsection{Client Operations}\label{sec:clientops}
To execute a command on the port knocking server, a remote user runs the client Python script, and the following operations are performed:
\begin{enumerate}
    \item The client supplied password is run through Argon2 to generate the associated key: \linebreak $\text{key}=\text{Argon2b}(\text{password})$.
    \item The client script pulls down and calculates the most recent random beacon from the blockchain API.
    \item The beacon and client supplied command are concatenated, and hashed with BLAKE2, using a keyed mode of operation: \\ $\text{knock}=\text{BLAKE2b}_{\text{key}}(\text{beacon}||\text{command})$
    \item The server's port number listening to the knocks is derived similarly to the the key, though instead the command is replaced with ``0''. The first two bytes of the resulting hash are then converted into an integer port number, this combined with the IP provide the destination to which the knock is sent.
\end{enumerate}

The key, beacon, and knock at $i$ are calculated as:

\begin{align*}
 \text{key} &= \text{Argon2i}\left(\text{password}\right), \\
   \text{beacon} &= \text{BLAKE2b}_{\text{key}}\left(\text{blockHeader} + \text{blockHeaderHash}\right),
    \text{knock}_{i} &= \text{BLAKE2b}_{\text{key}}\left(\text{beacon} || \text{command}_{i}\right)
\end{align*}

where $i$ represents the chosen command for the particular knock session (the server will listen for all possible values of $i$).

\subsubsection{Server Operations}
The structure for the server operations is largely similar to Prototype II: there are a number of interrelated tasks the server needs to perform concurrently, managed through Python's \texttt{multiprocessing} library:
\begin{itemize}
    \item The server periodically checks the blockchain API for new values, and stores harvested beacons in a \texttt{multiprocessing Manager} dictionary, to enable access by other processes. Once a new beacon is harvested, a boolean in the dictionary is flipped, signalling that the traffic sniffing process needs to update the valid knocks it listens to.
    \item Whenever a new beacon is found, for each command in the client profile a new valid knock string must be calculated (per Client Operation 3). Once generated, each knock string is sent to the listening process, using \texttt{scapy}. This listening process acts on the port number generated per client Operation.
    \item The listening process, checks each received UDP packet, of appropriate length, on the correct port, against the calculated valid knock values. If a match is found, the relevant command is sought and executed.
\end{itemize}

\subsubsection{Discussion}
The random port number derived from the beacon serves two purposes: firstly, it may make the job of an attacker listening on the wire slightly more difficult, as there is one less defining traffic characteristic to filter by; secondly, it improves usability, as the user no longer needs to memorise the port number to knock against. This change from Protocol II is made possible by the enormous difference in hash run-times.

\section{Evaluation}
\begin{figure*}[h]
\centering
\includegraphics[width=0.6\textwidth]{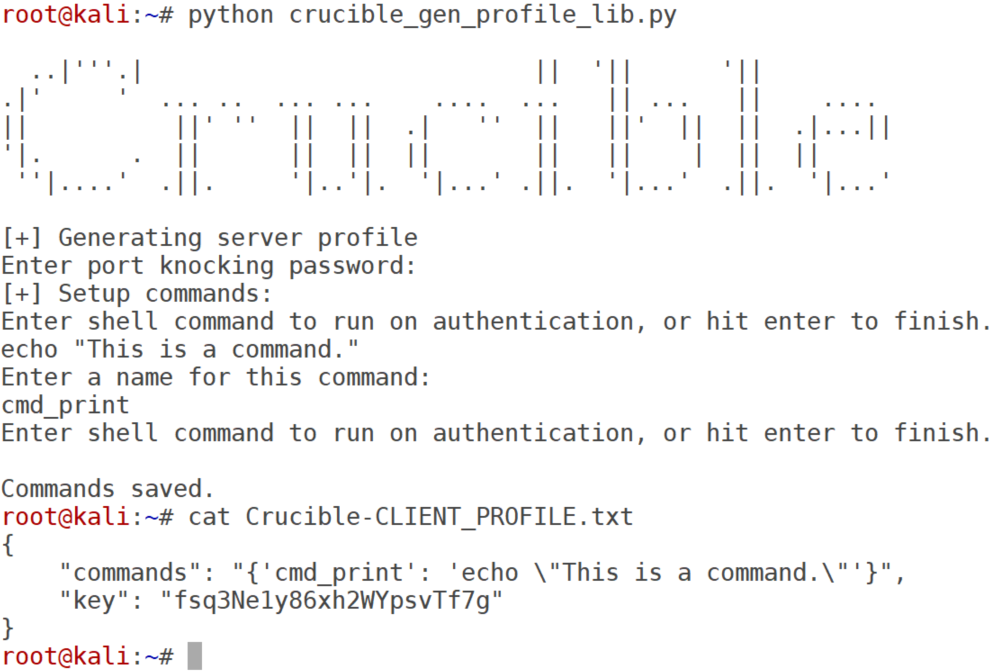}
\caption{Crucible setup: generating a client profile.}
\label{fig:gen}
\end{figure*}

\begin{comment}
The penultimate chapter of this work seeks to honestly appraise the proposed solution, Crucible. This is performed with the utmost efforts to remain impartial in assessment, to responsibly raise any identified weaknesses, and to avoid any misrepresentation of the work. The demonstration of Crucible will feature a gallery of screen captures from testing Crucible's abilities. This section is then followed by the bulk of the evaluative work in this chapter: Crucible is subjected to a range of attack scenarios, in order to assess its security. Before concluding, the chapter finishes with an overview of Crucible's merits and drawbacks.

\subsection{Crucible Demonstration}
\end{comment}
To begin using Crucible, the client must generate a profile and share this with the server. Figure \ref{fig:gen} shows this process, with the generated profile printed out. The key is derived from the user's password using Argon2, as per Section \ref{sec:Argon}.

\begin{figure*}[h]
\centering
\includegraphics[width=\textwidth]{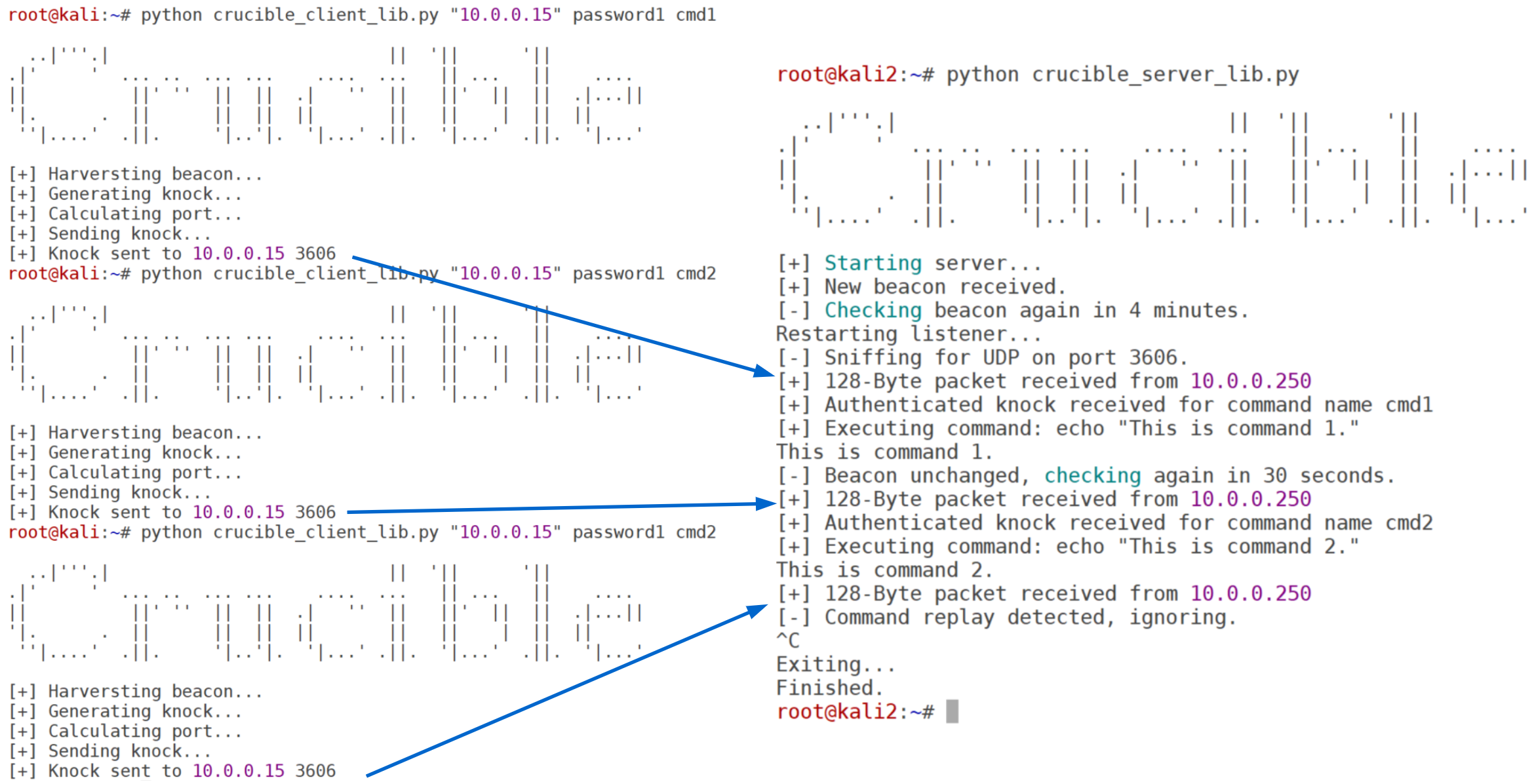}
\caption{Normal client and server operations of Crucible.}
\label{fig:6knock}
\end{figure*}

Figure \ref{fig:6knock} shows typical usage of Crucible. The client runs the Python script, passing in the IP, password and command parameters. There is also an interactive mode alternative for this operation. The client executes three commands, ``cmd1'', ``cmd2'' and ``cmd2''. As the latter command has already been executed by the server, it is ignored for replay protection. These commands will become available again when a new random beacon is sourced. 

\begin{figure*}[h]
\centering
\includegraphics[width=\textwidth]{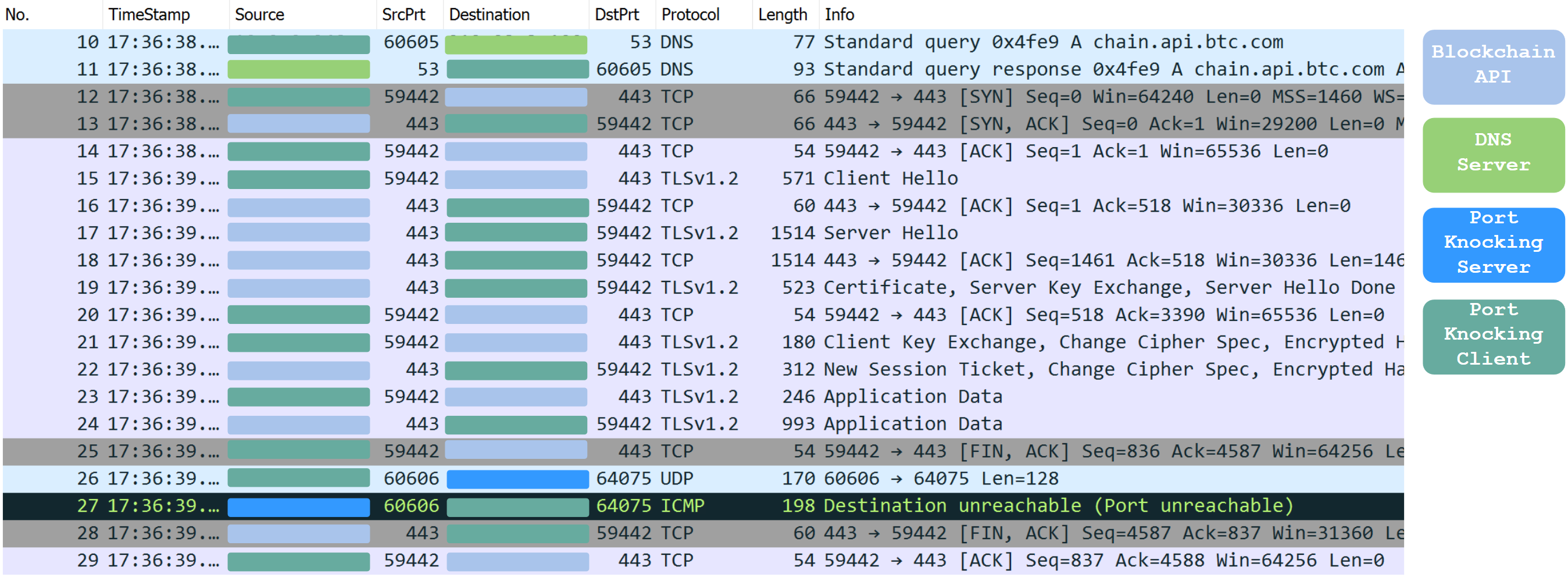}
\caption{Wireshark traffic capture of the client knocking process}
\label{fig:wire}
\begin{itemize}
    \itemsep0em
    \item Packets 10-12 show the client's DNS request for the Blockchain API's IP.
    \item Packets 12-24, and 28-29 show the random beacon retrieval over HTTPS.
    \item Packet 26 is the client knock.
    \item Packet 27 is a closed-port ICMP reply, per RFC 792 \cite{792}.
\end{itemize}
\end{figure*}

Figure \ref{fig:wire}, is a traffic capture of traffic emanating from the client machine as a result of port knocking with Crucible. It can be seen that Crucible does indeed only require a single packet for authentication and command authorisation. Further, the beacon data collected from the blockchain API is seen to be protected via HTTPS.

\begin{figure*}[h]
\centering
\includegraphics[width=0.6\textwidth]{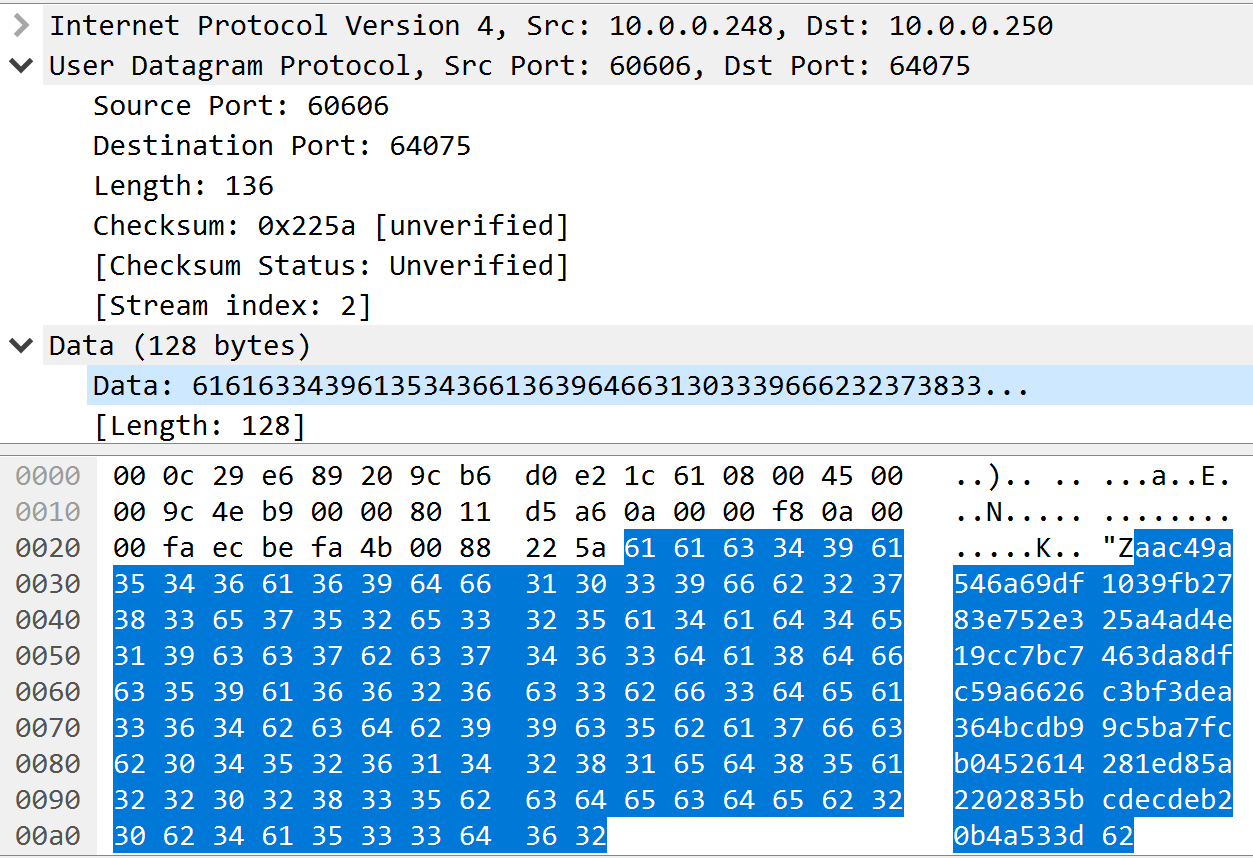}
\caption{Wireshark traffic capture of the client knock.}
\label{fig:wire-knock}
\end{figure*}

The knock itself, as seen in Figure \ref{fig:wire-knock} is a single 512-bit BLAKE2 hash sent via UDP, both the payload contents and the destination UDP ports are pseudo-randomly derived, the source port is ephemeral and chosen by the client OS. 

\begin{figure*}[h]
\centering
\includegraphics[width=\textwidth]{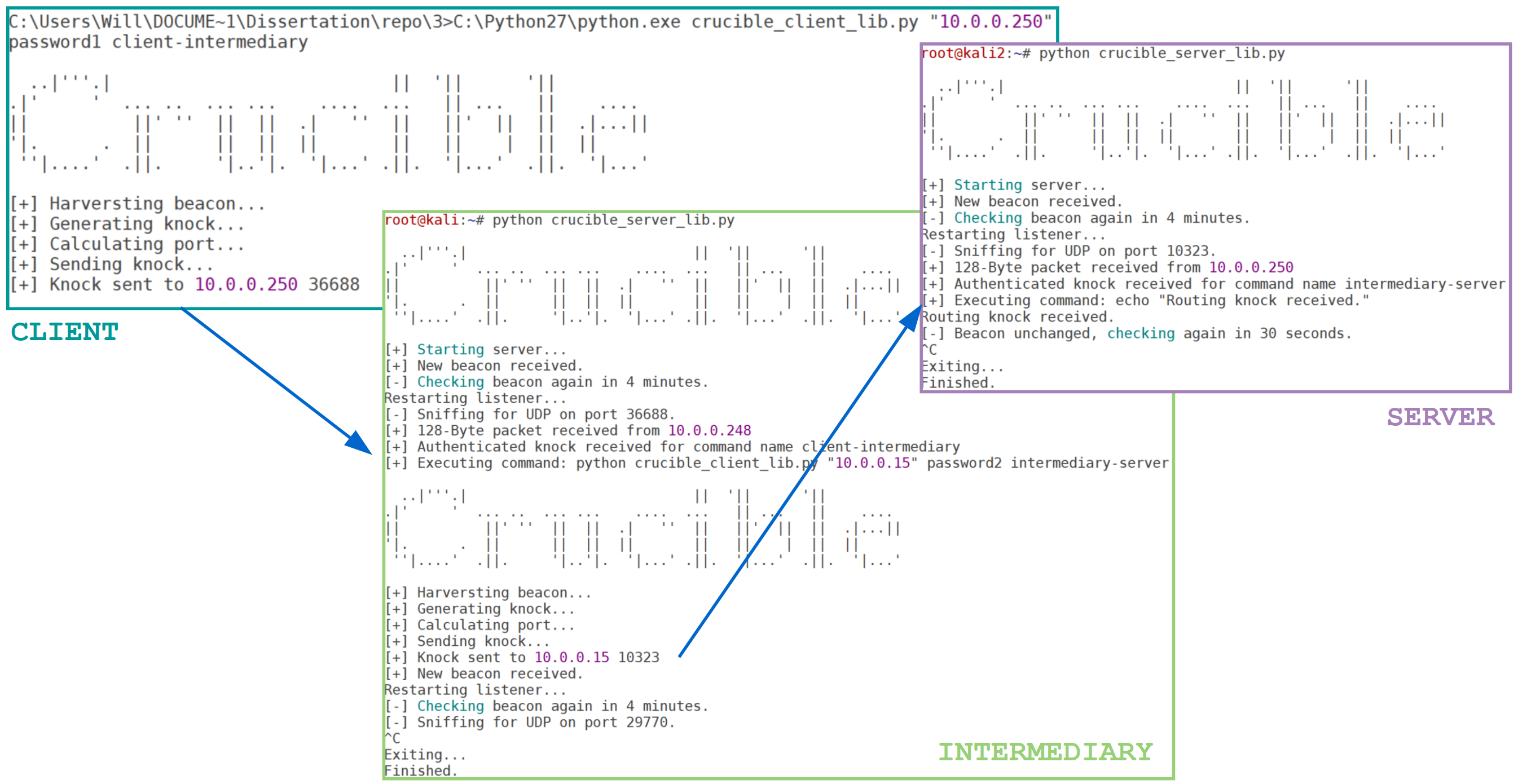}
\caption{Routing Crucible commands through intermediary hosts}
\label{fig:routing}
\end{figure*}

Figure \ref{fig:routing} demonstrates how Crucible avoids the NAT problems faced by port knocking solutions. This problem arises if a port knocking client attempts to authenticate with a server whose IP address has been translated, for example, by NAT\cite{1s}, proxies, or a VPN \cite{1t}. As a result, the knock packet sent by the client will not reach the intended destination. In such a scenario, Crucible should be installed on each intermediary host, and is set-up with a command to spawn a new client process, and knock on the next host in the chain, until the command reaches its final destination. 

% testing/demonstration
%   not very much 'testing'....
%   windows and linux
%   port scan - bit boring/small
%   delay/latency on wire - how?
%   cpu/ram usage on server

\subsection{Attack Modelling}
\begin{figure}[h]
\centering
\includegraphics[width=0.4\textwidth]{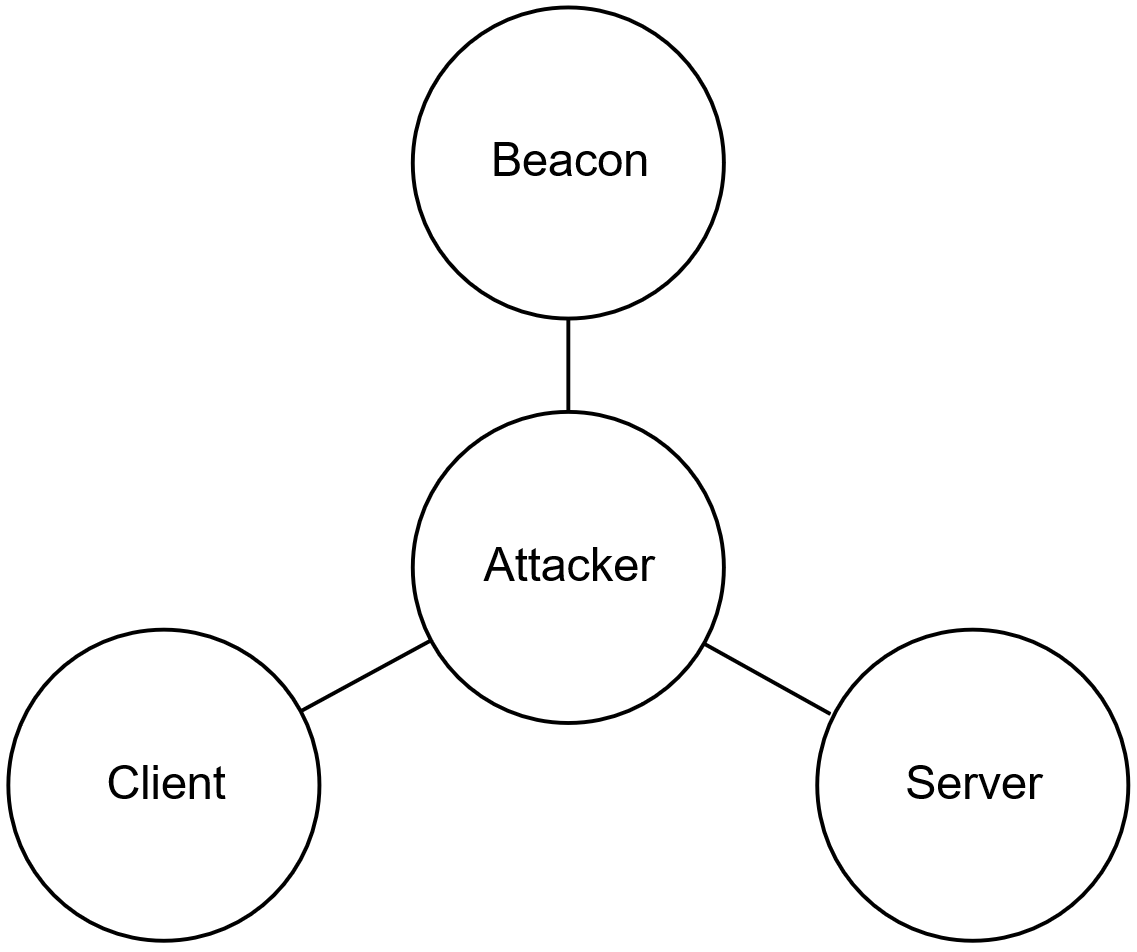}
\caption{Attacker positioning context.}
\label{fig:attack}
\end{figure}

In this section a range of potential attacks against the operations of Crucible are considered, exploring what capabilities an adversary may have, and what the ramifications of each attack are. \cite{serious} outlines the goals of attack modelling, which are paraphrased as follows:
\begin{itemize}
    \item To help set requirements for future protocol design.
    \item To provide users guidelines on whether a protocol will be safe to use in their environment.
    \item To provide clues for analysts keen to find weaknesses in the protocol, as part of the security process, so they can determine whether a given attack is valid. 
\end{itemize}
% TBC change `as follows'
Unless otherwise specified, the attack assumptions this section operates under are that the attacker is positioned per Figure \ref{fig:attack}. This simple diagram may seem pointless to include, but it helps formalise firstly that all networks Crucible operates across are considered untrusted. Secondly, the diagram highlights the potential extent of attacker capabilities under a worst-case scenario -- full traffic access, and interdiction. 

% extent of capabilities
% intermediaries

\subsubsection{Attacks on Identification Protocols}
\cite{2l} outlines a range of attacks that can be mounted against identification protocols, here they are reviewed in the context of Crucible:
\begin{itemize}
    \item ``impersonation'': an attacker impersonating the client will only be able to authenticate against the server with use of the password. The beacon service is open-access, and would not suffer an attacker imitating the client or server. Impersonating the server, for example in a man in the middle scenario, could grant an attacker access to knocks, and could prevent them reaching the legitimate server. From obtained knocks, there is arguably little an attacker could gain, as reversing the hash functions is intractable by design. If a knock value has already been issued, there is little to be gained from an attacker intercepting and relaying it themselves. Impersonation of the beacon, owing to HTTPS authentication via public key infrastructure, is assumed to be difficult.
    \item ``replay attack'': If a beacon value were replayed to the server, an attacker would have the opportunity to replay commands the client had already issued. If a beacon value was replayed to the client alone, the server would not accept that client's knocks. Both scenarios represent threats and illustrate the trust required in the beacon service, and its secure access. As the server is silent, only replay attacks seemingly sent from the client require further consideration, which are actively defended against per Figure \ref{fig:6knock}.
    \item ``interleaving attack: an impersonation or other deception involving selective combination of information from one or more previous or simultaneously ongoing protocol executions.'' Interleaving attacks don't appear applicable here as the protocol has little interactivity. Further information on such attacks can be found in \cite{6a}.
    \item `` reflection attack: an interleaving attack involving sending information from an ongoing protocol execution back to the originator of such information.'' See interleaving attacks.
    \item ``forced delay'': without a beacon, neither client nor server can operate port knocking functions. Delaying port knocks from client to server could make them stale, and therefore unacceptable.
    \item ``chosen-text attack: [...] an adversary strategically chooses challenges in an attempt to extract information about the claimant's long-term key.'' Impersonating the beacon, an attacker could try to send a message to recover the key through examination of the resulting knocks, though this would require subversion of the BLAKE2 hash function, which is both keyed, and invoked twice in the production of a knock value. It is unclear to the author how such an attack would be approached, and seemingly infeasible.
\end{itemize}

From the attacks discussed, most would require man-in-the-middle capabilities of an attacker. Impersonation of the beacon service for use in replay of a beacon, or chosen-beacon attacks, represent the only threats substantially different from an attacker with the capability to physically tamper with the network cable, causing delay or loss of service. This sort of impersonation could be carried out if an attacker had compromised the API's service, or if the requests made to the API via HTTPS were not sufficiently secured. If Bitcoin itself were manipulated, in order to influence beacon values, this should be considered in the same context as an attacker impersonating the beacon service. \cite{4f}, and \cite{mal} provide cost estimates for such an attack in the thousands of dollars. The chosen block outcome would further need to circumvent double-invocation of keyed BLAKE2.

\subsubsection{Online Attacks against the Listening Service}
The \texttt{scapy} interface for lower-level \texttt{libpcap} functionality could provide adversaries an attack vector for Crucible. Indeed, the \texttt{libpcap} is not without historical vulnerabilities \cite{6d} and applications of \texttt{libpcap}, such as Wireshark, have experienced vulnerabilities as a result of this \cite{6e}. As noted previously, Crucible requires root permissions (for its raw packet access), and an exploitation of \texttt{libpcap} or \texttt{scapy} could have dire consequences. The listening service itself could also be vulnerable to a denial of service attack, as is the case for IDPS devices, where an attacker may send large volumes of traffic, sometimes anomalous in nature, `` to attempt to exhaust a sensor's resources or cause it to crash'' \cite{6f}. For Crucible, with its fail closed stance, this would prevent client from authenticating until the service was restarted.

\subsubsection{Offline Attacks against Passwords}\label{sec:passwords}
In Crucible, the Argon2 hash is used as the key for BLAKE2, and this hash is retained on the server to authenticate clients. Should the server be compromised, and if its hash values are stolen and cracked (i.e. the passwords were discovered) an attacker could use these credentials to log into other client accounts, elsewhere. Password cracking is generally the process of inverting a given hash $H_i$, by discovering the password such that $H(password)=H_i$. This involves generating a great deal of passwords, hashing each, and comparing the results against the obtained hash to find a match.\\ 

If the password supplied to Argon2 is not of sufficient length, bruteforce attacks are made easier for an adversary, as there is less keyspace to search. Similarly, if the password is weakened by using predictable values (words, numbers, patterns etc.) then this makes dictionary attacks easier to mount. Further, the Argon2 hash is derived using a static salt value, meaning if future development added capability for multiple users, then the hashes derived from these passwords would be vulnerable to rainbow table attacks, whereby a single $H(password)$ can be tested against all of the server's user hashes. This could be solved by having the user remember a salt value (or username=salt), or by including a client state required for knocking, neither of which are preferable solutions. The key point, noted in (\ref{sec:Argon}), is that Argon's settings parametrise the difficulty for calculating a hash, making authentication {\it slightly} more slow for users, but dramatically increasing the cost to adversaries of mounting bruteforce or dictionary attacks.

\subsubsection{Attacks against Dependencies}\label{sec:depends}
Library and package dependencies are important in the context of secure protocol design, because external code vulnerabilities can result in exploitation of the prototype itself. As discussed earlier in Section \ref{sec:NSA}, a NIST elliptic curve cryptographic standard was allegedly backdoored. \cite{6g} is a more recent example of a Python module for handling SSH connections, which surreptitiously harvested and exfiltrated the user's SSH credentials. Javascript library BrowseAloud was recently compromised, resulting in infection of websites using the library with cryptojacking software; repurposing their machines as crypto-miners \cite{6h}. For transparency, Crucible's dependencies are included here in Table \ref{table:depends}, and their use should be properly reviewed before deployment in a production environment. Once deployed, the libraries need to be constantly updated, and periodically checked against vulnerability databases.\\

\begin{table}[h]
\caption{{{Dependencies in the Crucible prototype.}}}
\resizebox{\columnwidth}{!}{%
\begin{tabular}{|c|c|c|c|c|c|c|c|c|c|}
\hline
           & socket & getpass & Argon2 & pyblake2 & sys & multiprocessing & time & os & scapy \\ \hline
Client     & \checkmark      & \checkmark       & \checkmark      & \checkmark        & \checkmark   &                 &      &    &       \\ \hline
Server     &        &         &        & \checkmark        & \checkmark   & \checkmark               & \checkmark    & \checkmark  & \checkmark     \\ \hline
3\textsuperscript{rd} party? &        &         & \checkmark      & \checkmark        &     &                 &      &    & \checkmark     \\ \hline
\end{tabular}
}
\label{table:depends}
\end{table}

% BLAKE2 and Argon vs established standards...

\subsubsection{Reconnaissance and stealth}\label{sec:stealthrecon}
Reconnaissance encompasses the methods an adversary can deploy in information gathering at the start of a campaign. The level of {\it stealth} a port knocking implementation provides may determine whether or not it is detected by an attacker conducting reconnaissance, and therefore stealth can decrease the likelihood that a port knocking server is exploited. An attacker performing reconnaissance activities could benefit from the following information, all of which could be made possible through detection of port knocking:
\begin{itemize}
    \item Identification of a host as a client, server, or beacon service.
    \item Detection of a port knocking service on a server.
    \item Identification of the services hidden or protected by port knocking.
    \item Identification of the port knocking implementation i.e. Crucible.
\end{itemize}

There are a number of characteristics that could indicate port knocking from captured traffic between client, server and beacon. Assuming an attacker had access to Crucible's documentation and code, {\it passive methods} of traffic analysis could look for indications such as:

\begin{itemize}
    \item Periodic requests to the beacon service. These could be cross referenced with the beacon API to match a new block with a spike in HTTPS data transfer. If using the default API, an attacker could perform a reverse lookup to the API's URL and identify Crucible traffic this way.
    \item UDP packets with 512-bit payloads, where the destination port always differs. Other features of the knock packet may be identifiable, a number of techniques for passive fingerprinting of traffic are explored by \cite{1ac}. Higher amounts of UDP traffic than expected may contradict stealth goals \cite{1i}.
    \item Identifiable traffic resulting from a port knocking authorised command. For example, if the command was to open a port, an attacker could periodically diff open port scans of the server, and identify successful knocking traffic.
\end{itemize}

\begin{figure}[h]
\centering
\includegraphics[width=\columnwidth]{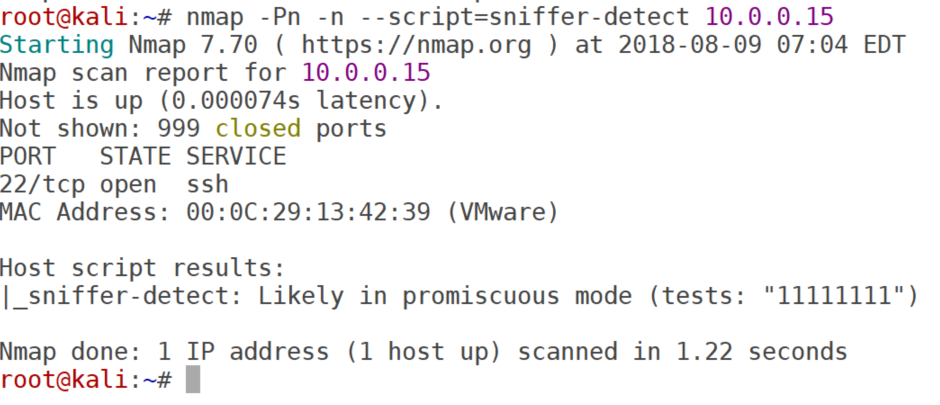}
\caption{Indication of a Crucible server: promiscuous mode NIC}
\label{nmap}
\end{figure}

\cite{6c} explores {\it active methods} an attacker can use to identify use of promiscuous mode by a NIC on the local network, which is active when Crucible is running. One such method involves crafting an ARP packet, which would normally be rejected by the NIC, however in promiscuous mode the host issues no such reply. As seen in Figure \ref{nmap}, the Nmap \texttt{sniffer-detect} (Nmap is a popular network sniffing tool) script performs this and other checks, and may allow an attacker to distinguish between Crucible and other port knocking solutions. 
%figure/label not working

Greater discussion on stealth aspects of port knocking is reviewed in \cite{1t}, \cite{1k} and \cite{1l}. In comparison with other solutions examined in Section \ref{sec:mechanics} on port knocking mechanics, Crucible has the following overall advantages and disadvantages in stealth:
\begin{itemize}
    \item With only a single packet sent to the server for the knock, proportionally less traffic on the wire is attributable to Crucible than other solutions with more client-server interactivity.
    \item No indication of the command executed on the server should be identifiable from the wire, as replay protection is enforced (no command will be re-issued), and as the command name is masked by keyed hashing. The definition of a keyed hash in Section \ref{hashes} explains why the command is not recoverable from traffic. No client identifiers such as IP addresses or usernames are recoverable from network traffic, nor are service identifiers such as the service destination port. Very little information is leaked: between client and server, the only traffic exchanged is a single datagram with a pseudo-random number as its payload.
    \item A single UDP datagram containing a random number may look suspicious in a given context, Crucible makes no effort to deploy steganography, or other methods, to appear innocuous to an attacker. 
    \item As a multi-party solution using a random beacon service, Crucible is more easily detected. 
\end{itemize}

%\cite{1z} due to the fact that most SPA payloads are seemingly random, it is difficult to obtain exact distinguishing characteristics, though an example is given (tracks default port, packet length)

\section{{{Conclusion}}}
Zero knowledge proofs were introduced for private, lightweight client-identification. Chaos-based cryptography was explored for the purpose of minimalist, dependency-free cryptographic hashing. Random beacons were used to secure replay protection while preserving a single-packet knock, and preventing attacker-chosen computation. This paper has explored novel combinations of these topics with regards to port knocking and has used these ideas to develop Crucible.

Crucible achieves command authorisation using a single packet between client and server, with a payload likely indistinguishable from a random number. Crucible's design is minimalist, secure and stateless. It is portable, and the user only needs to memorise an IP, a password, and a command name in order to authenticate. Crucible does not authenticate post-knock traffic (see \cite{1z}) and a single command can only be executed once within the period of a random beacon, though an unlimited number of unique commands can be run in this window. Lastly, trust must be placed in the random beacon service, which is a non-trivial consideration.

\section{Ethical Statement}
There are no potential conflicts of interest, and the research does not include human participants and/or animals. The work has been undertaken to accepted standards of ethics and of professional standards.

\par\bigskip
\section*{{{Appendix}}}

\rule{\columnwidth}{0.4pt}
{\bf Protocol 1 } Schnorr Identification Protocol\\[.01\normalbaselineskip]
\rule{\columnwidth}{0.4pt}
\textit{Protocol setup} 
\begin{itemize}
    \item Let $p$ and $q$ be two large primes where $p-1$ is a multiple of $q$. Let $G_{q}$ denote the subgroup of $\mathbb{Z}_{p}^{*}$ of prime order $q$, and $g$ be a generator for the subgroup.
    \item Let $a$ be the private exponent chosen uniformly at random from $[0, q-1]$.Let $A=g^a \mod{p} $.
    \item Ahead of execution, both prover and verifier know $\left(p, q, g, G, A \right)$.
\end{itemize}
\textit{Protocol actions}
\begin{enumerate}
\item Prover chooses a number $v$ uniformly at random from $[0, q-1]$ and computes $V=g^v \mod{p}$ and sends this to the Verifier.
\item Verifier chooses a challenge $c$ uniformly at random from $[0, 2^{t-1}]$, where t is the bit length of the challenge, and sends this to the prover.
\item Prover computes $r=v-ac \mod{q}$ and sends this to the Verifier.
\item Verifier ensures:
\begin{enumerate}
\item $A$ is within $[2, p-1]$
\item $A^q = 1 \mod{p}$
\item $V = g^r  A^c \mod{p}$
\item[]
\item[]
\end{enumerate}
\item \textit{Protocol messages}
\begin{center}\begin{tabular}{lcr}
Prover $ \rightarrow $ Verifier:&     $V=g^v \mod{p}$ & (1)\\
Prover $ \leftarrow $ Verifier:&  $c$ & (2)\\
Prover $ \rightarrow $ Verifier:&     $r=v-ac \mod{q}$ & (3)\\
\end{tabular}\end{center}
\end{enumerate}
\textit{Notes} \\
The protocol proves knowledge of the secret exponent $a$ without revealing any information about it. Setup parameters may be chosen as per DSA choices. Checks 4(a) and (b) are to avoid invalid public keys. Check 4(c), since: \\ $g^r  A^c = (g^{v-ac})(g^{a})^{c} = g^{(v-ac+ac)}=g^v = V  \mod{p}$. This protocol is referenced from \cite{8235}, Section 2.2.

\rule{\columnwidth}{0.4pt}

\rule{\columnwidth}{0.4pt}
{\bf Protocol 2 } Schnorr Non-interactive Zero-Knowledge Proof\\[.01\normalbaselineskip]
\rule{\columnwidth}{0.4pt}
\textit{Protocol setup} 
\begin{itemize}
    \item Let $p$ and $q$ be two large primes where $p-1$ is a multiple of $q$. Let $G_{q}$ denote the subgroup of $\mathbb{Z}_{p}^{*}$ of prime order $q$, and $g$ be a generator for the subgroup.
    \item Let $a$ be the private exponent chosen uniformly at random from $[0, q-1]$. Let $A=g^a \mod{p}$ be public key associated with $a$.
    \item Let $H$ be a secure cryptographic hash function, {\it UserID} a unique identifier for the Prover, and {\it OtherInfo} optional data.  The bit length of the hash output should be at least equal to that of the order $q$ of the considered subgroup.
    \item Ahead of execution, both prover and verifier know $\left(p, q, g, G, A, \text{H}, \text{UserID} \right)$.
\end{itemize}

\textit{Protocol actions}
\begin{enumerate}
\item Prover chooses a number $v$ uniformly at random from $[0, q-1]$ and computes the following:
\begin{enumerate}
    \item $V=g^v \mod{p}$.
    \item $c=\text{H} \left( g \| V \| A \| \text{UserID} \| \text{OtherInfo} \right)$
    \item  $r=v-ac \mod{q}$.
\end{enumerate}
\item Prover sends $\left( \text{UserID}, \text{OtherInfo}, c, r \right)$ to Verifier.
\item Verifier uses the provided UserID to lookup $A$.
\item Verifier computes $V = g^r A^c$.
\item Verifier ensures $c \stackrel{?}{=} \text{H} \left( g \| V \| A \| \text{UserID} \| \text{OtherInfo} \right)$.
\end{enumerate}

\rule{\columnwidth}{0.7px}\vspace{1mm}
{\bf Protocol 3 } Absolute Value Chaos-based Cryptographic Hash Function\\[.01\normalbaselineskip]
\rule{\columnwidth}{0.7px}\vspace{1mm}
%\begin{enumerate}[I.]
\textit{Inputs} 
\begin{itemize}
    \item A message string $M$ of arbitrary length.
    \item A key $K$ used as the initial value of the chaotic map. Chosen as a long floating number of 128-bit precision, and associated keyspace.
    \item A fixed number of map iterations $i$. %to apply the chaotic map for.
    \item A chosen range for the $\alpha$ coefficient to introduce chaos into the absolute value map, as seen in Equation \eqref{eq2}. 
\end{itemize}
\textit{Process}
\begin{enumerate}
\item The message is padded with the ASCII character '0' ASCII appended to the suffix, until the message length in bytes is a multiple of 8.
\item The message $M$ is split into an $L$-sized array of bytes $\omega_{i}$ for $i=1 \dots L$. Each byte element uses the ASCII integer value for the associated string character in $M$.
\item With key $K$ substituting for $\omega_{0}$ as the initialisation value, each $\omega_{i}$ is iterated through the chaotic map $i$-times: $x_{n+1} = 1-ABS((0.0015 \omega_{m} + 1.8) x_{n})$
\item This is repeated for the next $\omega_{m+1}$, using the previous $x_{i}$ as the initialisation value $\omega_{0}$, until each byte of the original message has been combined into $x_{L}$ the last value.
\item The final value produced by the chaotic map $x_{L}$ is normalised into a long number in the range $0 \leq H(M) < 2^{128}$ as the message digest.
\end{enumerate}

\textit{Notes}  
 In step (3) the $\alpha$ value is equal to $0.0015 \omega_{i} + 1.8$, this is because each $\omega_{i}$ is an integer ASCII value between 32 and 126, so these values are normalised onto the chosen range of $1.8 \leq \alpha < 2$ for the required chaotic behaviour. This protocol is referenced from \cite{3k} with modification to the $\alpha$ parameter.

\rule{\columnwidth}{0.7px}\vspace{1mm}
{{\bf Protocol 4 } Bitcoin Random Beacon\\[.01\normalbaselineskip]}
\rule{\columnwidth}{0.7px}\vspace{1mm}

\textit{Inputs} 
\begin{itemize}
\item Secure cryptographic hash function $H(m,k)$, here the chaos-based keyed hash function from earlier is used, with key $k$ for message $m$.
\item Block header $B_{H}$ and block header hash $B_{HH}$ of the Bitcoin network's most recent block, polled from a Bitcoin tracking website (a blockchain API) by through a HTTPS request.
\end{itemize}

\textit{Process}
\begin{enumerate}
\item Pull down the block header $B_{H}$ and the hash of the block header $B_{HH}$ from the website.
\item Calculate the binary OR of the header and the block header hash, $b=B_{H} + B_{HH}$
\item Output $H(b,k)$
\end{enumerate}

\textit{Notes}  \\
This protocol is referenced from the implementation by \cite{4j} of the Bitcoin Random Beacon proposed in \cite{4f}, with the following modifications: \\
Check is performed to validate the block header hash received from the website, this is to avoid importing libraries for calculating SHA hashes used in Bitcoin. \\
\cite{4j} chose a HMAC construction using SHA-256, instead the chaotic keyed hash from \ref{hash} is used.
\
%\item Output is fixed to the hash function's digest length.
The block header hash is included to strengthen against malicious miners aiming to influence the block header. Even if the header were tampered with, the resulting hash value of this header remains unpredictable.

\rule{\columnwidth}{0.7px}\vspace{1mm}

\bibliographystyle{IEEEtran}
\bibliography{main}
\end{document}